\documentclass[AMA,Times1COL]{WileyNJDv5}

\usepackage{amsmath}
\usepackage{amssymb}
\usepackage[many]{tcolorbox} 
\usepackage{float}
\usepackage{tabularx}
\usepackage{listings}

\tcbset{
    sharp corners,
    before skip = 0.2cm,    
    after skip = 0.5cm      
}                           
\newtcolorbox{calloutbox}{
    boxrule = 0pt  
}

\articletype{Article Type}%

\received{Date Month Year}
\revised{Date Month Year}
\accepted{Date Month Year}
\journal{Journal}
\volume{00}
\copyyear{2023}
\startpage{1}

\begin{document}

\title{The Estimands Framework and Causal Inference: Complementary not Competing Paradigms }

\author[1]{Thomas Drury}

\author[2]{Jonathan W. Bartlett}

\author[3]{David Wright}

\author[4]{Oliver N. Keene}

\authormark{Drury \textsc{et al.}}
\titlemark{The Estimands Framework and Causal Inference: Complementary not Competing Paradigms}

\address[1]{\orgname{GSK}, \orgaddress{\country{UK}}}

\address[2]{\orgname{The London School of Hygiene and Tropical Medicine}, \orgaddress{\country{UK}}}

\address[3]{\orgdiv{Statistical Innovation Department}, \orgname{AstraZeneca}, \country{UK}}

\address[4]{\orgname{KeenONStatistics}, \country{UK}}

\corres{Corresponding author: Thomas Drury \email{thomas.a.drury@gsk.com}}


\abstract[Abstract]{The creation of the ICH E9 (R1) estimands framework has led to more precise specification of the treatment effects of interest in the design and statistical analysis of clinical trials. However, it is unclear how the new framework relates to causal inference, as both approaches appear to define what is being estimated and have a quantity labelled an estimand. Using illustrative examples, we show that both approaches can be used to define a population-based summary of an effect on an outcome for a specified population and highlight the similarities and differences between these approaches.
We demonstrate that the ICH E9 (R1) estimand framework offers a descriptive, structured approach that is more accessible to non-mathematicians, facilitating clearer communication of trial objectives and results. We then contrast this with the causal inference framework, which provides a mathematically precise definition of an estimand, and allows the explicit articulation of assumptions through tools such as causal graphs. 
Despite these differences, the two paradigms should be viewed as complementary rather than competing. The combined use of both approaches enhances the ability to communicate what is being estimated.  We encourage those familiar with one framework to appreciate the concepts of the other to strengthen the robustness and clarity of clinical trial design, analysis, and interpretation.
}


\maketitle

\renewcommand\thefootnote{\fnsymbol{footnote}}
\setcounter{footnote}{1}

\section{Introduction} \label{sec:introduction}
There has been widespread adoption of the estimands framework by the pharmaceutical industry following the publication of ICH E9 (R1) \citep{ICH-E9-R1, Mallinckrodt2020, Fletcher2022}. There has also been some use of the framework in academia \citep{Cro2022-he}. The purpose of the framework is to provide context and clarity on what is being estimated in a clinical trial by formally defining a quantity labeled an estimand that has five components: Population, Treatments, Endpoint, Summary Measure, and Intercurrent Events (with strategies chosen to address them). The framework was designed to provide a common language for different stakeholders to discuss and define effects of interest in clinical trials \citep{ICH-E9-R1-Training}. It was also, in part, an acknowledgment that to gain a full understanding of an effect targeted in randomized clinical trials (RCTs), we often need to consider how post baseline events impact the outcomes of interest.

Although the framework is a positive step forward in the design, analysis and reporting of clinical trials, a potentially confusing factor is that the term “estimand” has been used to describe a similar idea in the Causal Inference literature for many years \citep{Hernan2024}. A further complication is that what is labelled an estimand in each approach appears to partially overlap, and both relate to effects on patient outcomes under different treatment conditions.

In this paper we review how ICH E9 (R1) and the causal inference literature define what effect is being estimated in the context of a randomized trial. We discuss the origins of the two approaches, how they relate, and provide examples illustrating how each can be used to define the effect of interest. Although there are pros and cons to both approaches, they share the overall aim of providing clarity on what is being estimated and we suggest that it may be beneficial to use aspects from both when defining the treatment effect targeted in clinical trials. We also show that, although not explicitly referenced, the ICH E9 (R1) estimands framework relies on concepts developed as part of causal inference methodology. Finally, we discuss our interpretation of some more ambiguous aspects of ICH E9 (R1) which are not clear and comment on some recent opinions from the causal inference literature about the utility of the estimand framework.

This paper is organised as follows: in section 2 we outline some examples of clinical trials describing potential effects that could be targeted, section 3 gives a brief history and outline of the ICH E9 (R1) estimands framework, section 4 reviews relevant aspects of causal inference. Section 5 illustrates how the example effects from section 2 could be formulated using both approaches. Section 6 contrasts and discusses several aspects of both the estimands framework and causal inference and finally we provide a conclusion in section 7.

\section{Example Clinical Trials} \label{sec:example_clinical_trials}
Randomised clinical trials provide a rigorous basis for comparing groups of patients assigned different pre-specified treatments. With complete follow-up, randomization guarantees that the observed treatment effect on the outcome is, aside from random error, due to the treatment each individual was assigned. However, this effect of assignment is a combination of the effect of taking the assigned treatment plus any subsequent effects from other events that occur after assignment, for instance, treatment discontinuation or use of additional (or rescue) medications. In many clinical trials, this effect of assignment may not be the most relevant clinical question of interest to all trial stakeholders and other questions are often important, including the effects attributable to specific treatments (or combinations of treatments).

Events such as treatment discontinuation or use of additional medication can affect the value and interpretation of the outcome measured. For some clinical questions of interest, this can lead to difficulty using the recorded outcome measures directly to address the research questions in the trial and it is typically necessary to provide more information about how these events are reflected in the effects being targeted.

We present three examples based on real clinical trials that exhibit some of the issues discussed above. We do not discuss the trials in detail but use some aspects as motivation to aid the discussion of the similarities and differences between the causal inference and ICH E9 (R1) approaches to defining what is to be estimated.

\subsection{Diabetes}
The PIONEER 1 trial compared oral semaglutide to placebo in adult patients with type 2 diabetes, with a primary endpoint of change from baseline HbA1c \citep{Aroda2019-ie}. During the trial, patients could discontinue assigned treatment and/or receive rescue medication if they did not maintain acceptable glycaemic control, both of which would likely change their HbA1c.  Overall, more placebo patients needed rescue medication compared to semaglutide, creating a differential impact between the groups.  Therefore, comparing change in HbA1c values at the end of the trial considers the combined effect of the \textit{semaglutide taken plus any rescue taken} and does not address the treatment effect caused by or attributable to \textit{taking only semaglutide itself for the duration of the trial versus only placebo for the duration of the trial}.

\subsection{Nasal Polyps}
The SYNAPSE trial compared mepolizumab and placebo in chronic rhinosinusitis with nasal polyps after 52 weeks \citep{Han2021-aj}. The co-primary endpoints were nasal polyps score and nasal blockage score at the end of the trial. A post baseline event was surgery for nasal polyps and treatment with mepolizumab was intended to reduce the number of patients having to undergo surgery. 
Although surgery is likely to improve subsequent nasal scores, it represents a failure of randomised treatment. A key question is how to reflect this adverse outcome of nasal surgery in the comparison of treatments.  Comparing nasal polyps scores without regard to surgery would ignore this important outcome and not address the causal effect of treatment on the overall clinical outcome.

\subsection{COVID-19}
In a COVID-19 vaccine trial \citep{Baden2021-zm} participants at high risk for SARS-CoV-2 infection or its complications were randomly assigned to receive two intramuscular injections of the mRNA-1273 vaccine or placebo 28 days apart. The primary end point was prevention of COVID-19 illness (defined as onset at least 14 days after the second injection) in participants who had not previously been infected with SARS-CoV-2. Of the 30,351 participants who received at least one dose of either the active vaccine or placebo, 941 participants were excluded from the primary analysis, mainly because of failure to take the second dose or not taking the second dose in the appropriate time window. Clinical interest centred on answering a causal question of the treatment effect of receiving both doses in the appropriate time window, versus placebo.

\section{The ICH Estimand Framework} \label{sec:ich_Estimands} 
The impetus for developing the estimands framework and revising ICH E9 originated from the seminal 2010 report by the US National Research Council (NRC) on the prevention and treatment of missing data in clinical trials \citep{NRC-Report}. The report raised issues with the standard practice at the time for the design and analysis of clinical trials in regard to how missing data were handled.  Importantly the report emphasised the role of the trial estimand, stating that \textit{“Estimation of the primary (causal) estimand, with an appropriate estimate of uncertainty, is the main goal of a clinical trial”}. In addition, there was a continuing concern that “per-protocol” type analysis which only includes the patients who complied fully with the trial protocol risked creating groups that were no longer similar and may not reflect a valid causal comparison \citep{Ranganathan2016-jy, Hernan2012-jx, Lachin2000-rl, Ellenberg1996-bv}.  As a result of these concerns, regulators led a working group to develop the estimands framework to be included as a revision to ICH E9.     
ICH E9 (R1) was finalised in 2019 and we only give an outline for the purposes of our discussions. In brief, the guidance defines a process that starts by determination of the trial objective and associated clinical question of interest. Once this has been agreed, a population-based quantity called “an estimand” (which we label ICH estimand) should be defined to address the clinical question of interest. Once this ICH estimand has been defined, an appropriate estimator can be selected. \\

\noindent \textbf{The guidance defines an ICH estimand to have five components:} 
\begin{itemize}
  \item Population –- the group of patients the treatment(s) are designed to benefit.
  \item Variable –- the outcome measure used to assess the benefit for each patient.
  \item Treatments –- the specific treatment(s) assigned to each group of patients and any comparisons of groups to be made.
  \item Summary  –- the type of statistic used to summarise the variable of interest.
  \item Intercurrent Events (IEs) and strategies –- the post baseline events that could impact the value or interpretation of the variable measured and how these events are accounted for if they occur.
\end{itemize}

The framework specifies that \textbf{all five components are required to define “the estimand”} and taken together, they provide enough context to fully understand what effect is being estimated. The guidance defines five basic strategies to address IEs. These strategies provide clarity about how the occurrence of IEs are reflected in the estimand being defined. Again, in brief the strategies can be understood as:

\begin{enumerate}
  \item Treatment Policy –- Any effects resulting from the occurrence of IEs are included in the overall effect of interest. If this strategy is applied to all the identified IEs, then this corresponds to the effect of treatment assignment.
  \item Hypothetical –- Any effects resulting from the occurrence of the IE are removed from the effect of interest. This targets an effect in the absence of these IEs occurring.
  \item Composite –- The occurrence of the IE is considered an informative but (generally) poor outcome for the patient and the outcome variable’s definition is somehow re-defined so its value changes if the IE occurs. 
  \item While on Treatment –- The focus of interest is the variable measure until the particular IE occurs. The label of this strategy is arguably ambiguous and in general it can be thought of as “while the IE has not occurred”.
  \item Principal Stratum –- generally, this targets an effect in a group of patients that would (or would not) have had the IE if they had been assigned to either (or any) of the treatments in the trial. This attempts to consider effects in a reduced group of patients where the occurrence (or not) of the IE does not depend on their treatment assignment.
\end{enumerate}

Most ICH estimands defined in later stage trials (Phase IIb and III) will have two or more IEs. The most frequent IEs in general are discontinuation of assigned treatment and/or use of additional (sometimes rescue) medication. Death may be an important IE in trials where it is not the outcome, but depending on the disease may occur infrequently. In trials measuring time to a particular type of event, failures from competing events, or death, may be IEs. 

The text used to define an ICH estimand has evolved in two basic forms: a table listing each of the five components explicitly, or a paragraph that incorporates each of the five components \citep{Lynggaard2022-eu}. The details for the IE strategies are also either explicitly stated, or alternatively incorporated into the definitions of treatment, population, or variable attributes. We provide both a table and paragraph description for all ICH Estimands in this paper. The ICH estimand framework itself has very little mathematical content and relies on text descriptions to communicate concepts and scientific research questions. 

ICH E9 (R1) provided minimal guidance on estimation but did emphasise the importance of assessing the sensitivity of conclusions to statistical assumptions. Regression modelling and missing data approaches such as multiple imputation and inverse probability weighting may be useful but given the complex definitions of what effect an ICH estimand targets, estimation is an evolving field with many opportunities for novel methods \citep{Roger2019-gj,Polverejan2020-qy,Guizzaro2021-px,Hartley2022-ej,Wolbers2022-nl,Wang2023-tq,Noci2022-uo,Drury2024-jy,Bell2024-yt}. In particular, some estimators from causal inference are being explored \citep{Olarte_Parra2023-th}.

\section{Estimands in Causal Inference} \label{sec:causal_estimands}
While there are actually multiple causal inference frameworks \citep{Rubin2005, Hernan2024, Pearl2010}, they all share the aims of providing a language to clearly define causal quantities of interest and to spell out the assumptions under which these quantities would be correctly estimated. Within the causal inference literature, the potential outcomes (counter-factual) framework \citep{Rubin2005} is arguably the most popular and is the methodology we use for comparison with the ICH estimands framework. This framework was devised by Rubin and expanded by Hernan and Robins \citep{Hernan2024}, among others \citep{Pearl2010}.  Pearl (2022), Robins (2022) and Rubin (2022) have provided a thorough discussion of the history of causal inference \citep{Pearl2022, Robins2022, Rubin2022}. In order to discuss estimands defined in causal inference, we give a \underline{brief} description of the potential outcomes paradigm.

\subsection{Potential Outcomes Framework}
We consider the simple setting with individual patients that have a particular illness, and we wish to assess any benefit they might receive from a new drug in comparison to a control. In order to measure the benefit, we use an outcome measure ${Y}$ and    consider the existence of two potential outcomes for $Y$ in the theoretical scenarios where the individual is assigned the control or new drug. For ease of notation, we define the actual treatment assignment variable for an individual as ${A=0,1}$ and a particular (theoretical) treatment assignment as a superscript $a=0,1$ above the outcome. We then have two potential outcomes $Y^{\:(a\:=\:0)}$ and $Y^{\:(a\:=\:1)}$ which correspond to the outcomes the individual would experience if they were assigned the control or new drug respectively. The theoretical causal effect of the new drug on the individual could then be considered as the difference in these potential outcomes $Y^{\:(a\:=\:1)}-Y^{\:(a\:=\:0)}$. 

The potential outcomes framework relies on imagined worlds where different treatments are assigned to the same individual, but in reality, only one world is observed where the individual is assigned a single treatment. This means we cannot directly estimate individual causal effects. However, under certain conditions, we can estimate average effects in the population of individuals such as the difference in expected potential outcomes for individuals under each treatment $\mathbb{E}[Y^{\:(a\:=\:1)}]-\mathbb{E}[Y^{\:(a\:=\:0)}]$ which is equal to the average of the individual level causal effects. The basic conditions that need to hold in order to estimate this are referred to in the causal inference literature as: Consistency, Positivity and Exchangeability. We only give a very basic idea of these assumptions for the purpose of this work. For a thorough discussion see “What If” by Hernan and Robins \citep{Hernan2024}:

\begin{itemize}
  \item Consistency -– this means that the observed outcome under the treatment an individual is actually assigned corresponds to the individual’s potential outcome if assigned that treatment.
  \item Positivity –- this means that there is a non-zero chance that the individual could have received both (or any) of the treatments being compared.
  \item Exchangeability –- the chance of a patient being assigned each treatment is independent of the two potential outcomes.  
\end{itemize}

In randomized trials, these assumptions typically hold as randomization ensures a non-zero chance of being independently assigned to each treatment, so in addition to consistency, both positivity and exchangeability are satisfied.  Under these assumptions, the expected values of the potential outcomes are equal to the expected values for the groups of individuals assigned to each treatment in the trial. Specifically, $\mathbb{E}[Y^{\:(a=1)}] = \mathbb{E}[Y \:\vert\: A=1]$ (and similarly for control). This means the difference in expected potential outcomes $\mathbb{E}[Y^{\:(a=1)} - Y^{\:(a=1)}]$ can be estimated using $\mathbb{E}[ Y \:\vert\: A=1 ] - \mathbb{E}[ Y \:\vert\: A=0 ]$.  
	
In observational studies where assignment is not random by design, the positivity and exchangeability assumptions are often not satisfied. The exchangeability assumption can be weakened to a conditional version, which says that treatment assignment is independent of potential outcomes conditional on confounders. Estimation methods then must account for all the assumed confounders $ {W}$ using methods such as outcome regression models, or inverse probability weighting with propensity scores.

In more complicated settings, more complex causal ideas, notation, and assumptions are needed. As an example, in trials with longitudinal follow up, a patient’s treatment may change over time and there are different potential treatment sequences they could experience.  Expanding our previous set up to include a variable $ {D_j = 1,0}$ which corresponds to the situation of a patient receiving their treatment correctly at visit $ {j}$, we create a vector of indicators $ {\overline{D}=[D_1, \dotsc ,D_J]}$ that represent a treatment sequence. For example, with four visits, an imposed sequence of  $ {\overline{d}=[1,1,0,0]}$ would indicate a theoretical scenario where a patient receives their assigned treatment for the first two visits, but not the final two visits.

Although repeated follow up increases the complexity, for each potential treatment assignment $ {A}$ and sequence $ {\overline{D}}$ we are still able to define potential outcomes $ {Y^{(A, \:\overline{D})}}$ for a patient and consider these potential outcomes under specified (sometimes referred to as intervened) treatment assignments and sequences $ {Y^{(a, \:\overline{d})}}$. These types of treatment sequences are known as so-called "static regimes". More recent work in causal inference has expanded on this to "dynamic regimes" where the treatment of a patient can depend on time-varying follow up measures \citep{Tsiatis2019}. 

A tool developed in the causal inference literature to communicate the assumed structural relationships between the outcome, assignment process, and other factors are causal graphs. There are different types of causal graphs, developed for the different causal frameworks, but we consider the most popular which is the directed acyclic graph (DAG) \citep{Greenland1999}. In general, these graphs have arrows to indicate assumed causal effects which help practitioners visualize and communicate how different variables affect each other. We include a simple example of a DAG in Figure 1 but for brevity, we do not illustrate DAGs further in this paper.

\begin{figure}[h] 
  \caption{A simple directed acyclic graphs showing one possible causal structure of a clinical trial where baseline factors $ {W}$ and treatment assignment at randomization $ {A}$ have a causal effect on the outcome $ {Y}$ but also have a causal effect on patient discontinuation from randomized treatment $ {D}$ which also impacts the outcome.}
  \centering
  \includegraphics[width=0.4\textwidth]{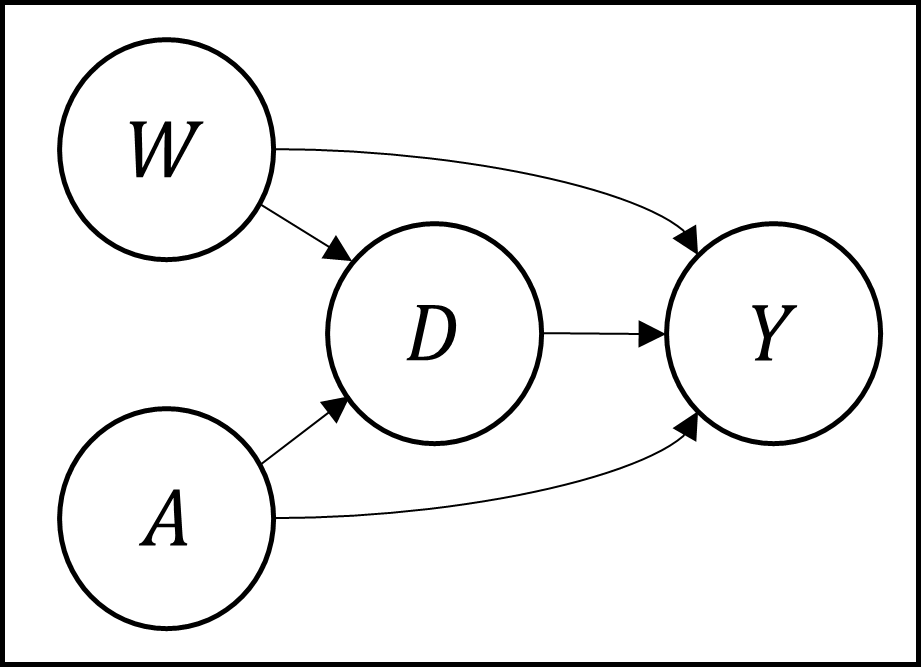}
\end{figure}

A wide variety of estimation methods have been developed within the causal inference literature, but they can generally be grouped as matching, inverse probability weighting and the so called “G-methods”, which includes G-formula and G-estimation. All these methods usually involve some form of (semi) parametric regression model. For a description of these methods see “What If” \citep{Hernan2024}. 

\subsection{Causal Estimands}
In the causal inference literature, an estimand (which we call a causal estimand) is \textbf{a mathematical expression to represent a summary comparison of potential outcomes under specified treatment conditions for a given population of individuals}. It represents a comparison of quantities that exist for a population and does not depend on what estimation method is used. The most basic example of a causal estimand is the difference in expected values for two treatments: $ {\mathbb{E}[Y^{(a=1)}] - \mathbb{E}[Y^{(a=0)}]}$, but other simple examples include the risk difference for two treatments: $ {\mathbb{P}(Y^{(a=1)}=1) - \mathbb{P}(Y^{(a=0)}=1)}$ or the risk ratio for two treatments $ {\mathbb{P}(Y^{(a=1)}=1) \:/\: \mathbb{P}(Y^{(a=0)}=1)}$. An example of a simple estimand in the repeated follow up setting discussed could be the difference in the expected values of outcome at the final follow-up between two treatments if they were correctly administered across all visits, specifically: ${\mathbb{E}[Y_J^{\:(a\:=\:1,\:\overline{d}\:=\:[1, \dotsc ,1]}] -\mathbb{E}[Y_J^{\:(a\:=\:0,\:\overline{d}\:=\:[1, \dotsc ,1])}]}$.

\section{Examples Revisited} \label{sec:examples_revisited}
We now revisit the examples discussed in section two and illustrate how these effects of interest could be specified in both the ICH estimand and causal inference frameworks. 

\subsection{Example 1 - Diabetes}
For this example, we consider a situation where the objective of PIONEER 1 might have been to answer these two different clinical questions:

\begin{calloutbox}
  \begin{enumerate}
     \item What is the effect of being assigned semaglutide vs placebo on HbA1c values after a year.
     \item What is the effect of semaglutide vs. placebo if patients were treated continuously for one year.
  \end{enumerate}
\end{calloutbox}

During the PIONEER 1 trial, a number of patients stopped taking the treatment they were assigned. A number also received rescue medication for glycaemic control. Both of these post baseline events are likely to impact the HbA1c efficacy measure.

\subsubsection{Estimands Framework}
In the estimands framework discontinuing assigned treatment and initiation of rescue medication would be considered IEs. These would then be included as the fifth component of the ICH estimand together with handling strategies that align with the clinical question of interest above.  We provide potential ICH estimands for each question below: \\

\noindent \textbf{ICH Estimand 1.1} \\
\textit{Difference in mean change from baseline HbA1c at week 26 for adult patients with type 2 diabetes assigned to semaglutide or placebo including any subsequent effects of discontinuation of assigned treatment or initiation of rescue medication.}

\begin{table}[hbt!]
 \begin{tabularx}{\textwidth}{l l} \hline  
   Population & Adult patients with type 2 diabetes \\ \hline  
   Treatment & Semaglutide or placebo, including any subsequent discontinuation or initiation of rescue medication \\ \hline  
   Variable & Change from baseline HbA1c at week 26 \\ \hline  
   Summary Measure & The difference in means \\ \hline  
   IEs (Handling) & Discontinuation of assigned treatment (Treatment Policy) \\
                  & Initiation of rescue medication (Treatment Policy)\\ \hline 
 \end{tabularx}
\end{table}

\noindent\textbf{ICH Estimand 1.2} \\
\textit{Difference in mean change from baseline HbA1c at week 26 for adult patients with type 2 diabetes between semaglutide and placebo if participants take their assigned treatment for the duration of the treatment period and do not use rescue medication.}
\begin{table}[hbt!]
 \begin{tabularx}{\textwidth}{l l} \hline  
   Population & Adult patients with type 2 diabetes \\ \hline  
   Treatment & Semaglutide or placebo throughout the full follow-up period \\ \hline  
   Variable & Change from baseline HbA1c at week 26 \\ \hline  
   Summary Measure & The difference in means \\ \hline  
   IEs (Handling) & Discontinuation of assigned treatment (Hypothetical) \\
                  & Initiation of rescue medication (Hypothetical) \\ \hline 
                  
 \end{tabularx}
\end{table}

\subsubsection{Causal Inference} 

In the causal inference approach, we could consider an individual and define their possible treatment assignment at randomization as ${A=0,1}$ representing placebo or semaglutide respectively. We then define their change from baseline HbA1c values as repeated measures ${Y_j}$ at visit ${j=1,…,J}$ with the final change from baseline HbA1c as the outcome of interest ${Y_J}$.  We also define ${D_j=0,1}$ as the discontinuation of assigned treatment at visit ${j}$ and create a vector over the trial visits as ${\overline{D}=[D_1,\dotsc ,D_J]}$. Similarly, we also define use of rescue therapy as ${R_j=0,1}$ and ${\overline{R}=[R_1, \dotsc ,R_J]}$. This allows the creation of potential outcomes of ${Y_J^{\:(A,\:\overline{D},\:\overline{R})}}$. Using this notation, we can then create a DAG to illustrate our assumptions about the causal relationships between the randomized assignment, treatment discontinuation, use of rescue, and the post-baseline HbA1c values. In some cases, it may also be necessary to consider other post baseline (time-varying) factors. Finally, we can specify the causal estimands that we are interested in via the following mathematical statements about the potential outcomes. \\

\noindent\textbf{Causal Estimand 1.1}
$$
{\mathbb{E}[Y_J^{\:(a\:=\:1,\:\overline{D}, \:\overline{R})}]-\mathbb{E}[Y_J^{\:(a\:=\:0,\:\overline{D}, \:\overline{R})}]=\mathbb{E}[Y_J^{\:(a\:=\:1)}]-\mathbb{E}[Y_J^{\:(a\:=\:0)}]}
$$

As $\overline{D}$ and $\overline{R}$ are not intervened on in this estimand we can simplify the estimand notation by removing them (last expression). This estimand is then the difference in expected values for a patient assigned to semaglutide vs placebo without regard for discontinuation or use of rescue medication. \\

\noindent\textbf{Causal Estimand 1.2}
$$
{\mathbb{E}[Y_J^{\:(a\:=\:1,\:\overline{d}\:=\:\textbf{0}, \:\overline{r}\:=\:\textbf{0})}]-\mathbb{E}[Y_J^{\:(a\:=\:0,\:\overline{d}\:=\:\textbf{0}, \:\overline{r}\:=\:\textbf{0})}]}
$$

Where ${\textbf{0}=[0,\dotsc,0]}$ corresponds to a ${1 \times J}$ vector of zeros. In words, this is targeting the difference in expected outcomes for a patient assigned to semaglutide vs placebo if they received their assigned treatment for the duration of the trial and do not initiate rescue medication. 

\subsection{Example 2 - Nasal Polyps}

For this example, we only concentrate on the first co-primary endpoint and consider a situation where part of the objective for the SYNAPSE trial was to answer the clinical question: \\

\begin{calloutbox}
  \begin{enumerate}
     \item What is the effect of being assigned mepolizumab vs placebo on nasal polyp score at the end of the trial where nasal polyp surgery is considered a poor outcome.
  \end{enumerate}
\end{calloutbox}

A key issue in the clinical question is how best to represent nasal polyp surgery in terms of the outcome measure nasal polyp score. As requiring surgery is a bad outcome for patients, it is informative about the efficacy provided by either assigned treatment. In this example one option to reflect surgery occurring would be to consider those patients as having the worst possible nasal polyp score.

\subsubsection{Estimands Framework}

In the estimand framework discontinuing assigned treatment and nasal polyp surgery would be considered IEs. These would then be included as the fifth component of the ICH estimand together with handling strategies that align well with the clinical question of interest above.  We provide a potential ICH estimand for the clinical question below: \\

\noindent\textbf{ICH Estimand 2.1} \\
\textit{Difference in mean nasal polyp Score at 52 weeks for adult patients with chronic rhinosinusitis assigned to mepolizumab or placebo including any subsequent effects of discontinuation of assigned treatment and considering nasal polyp surgery as a poor outcome reflected as the worst possible nasal polyp score.}

\begin{table}[hbt!]
 \begin{tabularx}{\textwidth}{l l} \hline  
   Population & Adult patients with chronic rhinosinusitis \\ \hline  
   Treatment & Mepolizumab or placebo, including any subsequent discontinuation  \\ \hline  
   Variable & Nasal polyp score at 52 weeks \\ \hline  
   Summary Measure & The difference in means \\ \hline  
   IEs (Handling) & Discontinuation of assigned treatment (Treatment Policy) \\
                  & Nasal polyp surgery (Composite where surgery is represented as the worst possible nasal polyp score) \\ \hline 
 \end{tabularx}
\end{table}

\subsubsection{Causal Inference}

In the causal inference approach, we use the same notation as the first example for outcome, assignment, visit and discontinuation of assigned treatment. We also define an individual having nasal polyp surgery by visit $j$ as $S_j=0,1$. This allows the creation of two sets of potential outcomes, one for the nasal score ${Y_J^{\:(A,\:\overline{D})}}$, and also another for surgery prior to visit $J$ as ${S_J^{\:(A,\:\overline{D})}}$. We could also create a DAG to communicate our assumptions about the causal relationships between randomized assignment, treatment discontinuation, surgery, and the post-baseline nasal polyp scores. Finally, with notation defined we can describe the following estimands. \\

\noindent\textbf{Causal Estimand 2.1} \\
For this estimand, there are two components to the effects of interest, first, any direct effect of assignment on the nasal polyp score and second, any indirect but informative effect of assignment on nasal polyp surgery. To reflect this, we construct a new outcome $Y_J^{\star}$ that is a mixture of nasal score potential outcomes for patients when no nasal polyp surgery performed, and the value chosen to reflect the poor outcome for patients where nasal polyp surgery would be performed. Again, as with causal estimand 1.1, there is no condition applied to $\overline{D}$ and therefore it is not included and the new outcome becomes:
$$
Y_J^{\star \: (A)} = {\mathbb{I}\left(  S_J^{\:(A)} = 0 \right)} \cdot {Y_J^{\:(A)}} \: + \: {\mathbb{I}\left(  S_J^{\:(A)} \neq 0 \right)} \cdot \theta  
$$
Where $\mathbb{I}\left(\cdot\right)$ are indicators of the condition contained in the brackets and $\theta$ is the value chosen for the individual to represent the poor outcome. In the SYNAPSE example, we consider a case when an exact value of worst possible nasal polyp score is defined as the value to represent a poor outcome in both arms ($\theta$ is set as a score of 4 in each nostril making a value of 8) and so our resulting estimand at the final timepoint $J$ is:
$$
\mathbb{E}[Y_J^{\star \: (a=1)}] \:-\: \mathbb{E}[Y_J^{\star \: (a=0)}] = \mathbb{E}\left[{\mathbb{I}\left(  S_J^{\:(a=1)} = 0 \right)} \cdot {Y_J^{\:(a=1)}} \: + \: {\mathbb{I}\left(  S_J^{\:(a=1)} \neq 0 \right)} \cdot 8 \right] \:-\: \mathbb{E}\left[{\mathbb{I}\left(  S_J^{\:(a=0)} = 0 \right)} \cdot {Y_J^{\:(a=0)}} \: + \: {\mathbb{I}\left(  S_J^{\:(a=0)} \neq 0 \right)} \cdot 8 \right]  
$$

\noindent{Applying the expectations leads to:}
\begin{multline*}
\mathbb{E}[Y_J^{\star \: (a\:=\:1)}] \:-\: \mathbb{E}[Y_J^{\star \: (a\:=\:0)}] = \left[ \mathbb{P}\left(S_J^{\:(a\:=\:1)} = 0 \right)  \cdot \mathbb{E}\left[Y_J^{\:(a\:=\:1)} \:\vert\: S_J^{\:(a\:=\:1)} = 0 \right] \: + \: \mathbb{P}\left(S_J^{\:(a\:=\:1)} \neq 0 \right) \cdot 8 \right] \\
- \left[ \mathbb{P}\left(S_J^{\:(a\:=\:0)} = 0 \right) \cdot \mathbb{E}\left[Y_J^{\:(a\:=\:0)} \:\vert\: S_J^{\:(a\:=\:0)} = 0 \right] \: + \: \mathbb{P}\left(S_J^{\:(a\:=\:0)} \neq 0 \right) \cdot 8 \right]  
\end{multline*}

In words, this estimand is the difference in expected combined nasal polyp score for a patient assigned to mepolizumab vs placebo without regard to treatment discontinuation. The expected combined nasal polyp score under each treatment is the weighted combination of the expected nasal polyp score among those not needing surgery and the worst possible nasal polyp score, with weights determined by the proportion of patients who need surgery under the assigned treatment. The weighting of the expected values builds in any causal difference in the probability of requiring surgery from being assigned to mepolizumab or placebo.

As discussed in Keene 2018 \citep{Keene2018}, the penalty value used for estimand 2.1 (defined using ICH or causal inference) is somewhat arbitrary. It also has the consequence of creating an effect that is specific to the $\theta$ value(s) used as the penalty term(s). This can make it difficult to compare these types of effects across trials if there is not general agreement on the penalty value.

Importantly, this example also shows that a clinical question of interest may require an estimand that uses different strategies for different IEs.

\subsection{Example 3 - COVID-19}

For this example, the primary endpoint was occurrence of symptomatic COVID-19 with onset at least 14 days after the second injection and we focus on the key issue that some participants failed to receive two doses of assigned treatment. (Occurrence of symptomatic COVID-19 within 14 days of randomisation represents another relevant issue that we will not discuss here).  

We consider a situation where the objective was to answer these two different clinical questions:

\begin{calloutbox}
  \begin{enumerate}
     \item What is the effect of being assigned mRNA-1273 vaccine compared to placebo on COVID-19 prevention if all participants were treated correctly.
     \item What is the effect of being assigned mRNA-1273 vaccine compared to placebo on COVID-19 prevention for the participants that would receive treatment correctly when assigned to either treatment.
  \end{enumerate}
\end{calloutbox}

This example illustrates the subtle difference between interest in an effect on all participants in a hypothetical setting where the IE would not occur for any participants, and an effect specifically for participants where the IE would not have occurred after being assigned either treatment. The former is an effect relating to the whole population, whereas the latter is an effect in a reduced set of participants from the original population.

Although a variety of approaches have been used to derive vaccine efficacy, we concentrate on using a relative risk based measure for vaccine efficacy.

\subsubsection{Estimands Framework}

In the estimand framework failure to receive two doses of assigned treatment in the correct time window would be considered an IE. This would then be included as the fifth component of an ICH estimand together with a handling strategy that aligns with the clinical questions of interest above.  We provide potential ICH estimands for each question below: \\

\noindent\textbf{ICH Estimand 3.1} \\
\textit{The percentage vaccine efficacy (derived from a risk reduction ratio) at 120 days for COVID-19 infection when assigned mRNA-1273 vaccine or placebo in adults with no known history of COVID-19 who are in circumstances that put them at risk of COVID-19 infection \textbf{assuming that they all received} two doses of their assigned treatment in the correct time window.}

\begin{table}[hbt!]
 \begin{tabularx}{\textwidth}{l l} \hline  
   Population & Adults with no known history of COVID-19 in circumstances that put them at risk of \\ & COVID-19 infection \\ \hline  
   Treatment & mRNA-1273 vaccine or placebo, two doses, in the correct time window \\ \hline  
   Variable & COVID-19 infection occurring more than 14 days after receiving treatment \\ \hline  
   Summary Measure & Vaccine Efficacy at 120 days calculated as percentage risk reduction ratio $(VE\%=100\cdot(1-RR)\%)$ \\ \hline  
   IEs (Handling) & Failure to receive two doses of assigned treatment in the correct time window (Hypothetical) \\ \hline 
 \end{tabularx}
\end{table}

\noindent\textbf{ICH Estimand 3.2} \\
\textit{The percentage vaccine efficacy (derived from a risk reduction ratio) at 120 days for COVID-19 infection when assigned mRNA-1273 vaccine or placebo in adults with no known history of COVID-19 who are in circumstances that put them at risk of COVID-19 infection \textbf{that would receive} two doses of assigned treatment in the correct time window if assigned to either treatment.}

\begin{table}[hbt!]
 \begin{tabularx}{\textwidth}{l l} \hline  
   Population & Adults with no known history of COVID-19 in circumstances that put them at risk of COVID-19 \\ & infection that would receive two doses of treatment assigned to them in the correct time window \\ \hline  
   Treatment & mRNA-1273 vaccine or placebo  \\ \hline  
   Variable & COVID-19 infection occurring more than 14 days after receiving treatment \\ \hline  
   Summary Measure & Vaccine Efficacy at 120 days calculated as percentage risk reduction ratio $(VE\%=100\cdot(1-RR)\%)$ \\ \hline  
   IEs (Handling) & Failure to receive two doses of assigned treatment in the correct time window (Principal Stratum) \\ \hline 
 \end{tabularx}
\end{table}

\subsubsection{Causal Inference}

In the causal inference approach, we use the same notation as the previous example for outcome, however we note COVID-19 infection is now a binary outcome. We also define a dosing indicator $C=0,1$ where $C=1$ corresponds to receiving the two doses of assigned treatment in the correct time window and this allows the creation of binary potential outcomes $Y^{\:(A, \: C)}$ which take the value 1 if the individual had an infection in the time window of 14 days after receiving treatment and the end of the trial. We could also create a DAG to communicate our assumptions about the casual relationships between the randomized assignment, treatment discontinuation, and COVID-19 infection. With notation defined we can describe the first causal estimand.\\

\noindent\textbf{Causal Estimand 3.1}
$$
{100 \cdot \left( 1 - \frac{\mathbb{E}[Y^{\:(a\:=\:1,\:c\:=\:1)}]}{\mathbb{E}[Y^{\:(a\:=\:0,\:c\:=\:1)}]} \right) }
$$
In words, this estimand is a percentage reduction of infection calculated using risk ratio of infection if participants were assigned mRNA-1273 vaccine or placebo assuming all participants received both doses of assigned medication in the correct time window.\\

\noindent\textbf{Causal Estimand 3.2} \\
For this estimand we extend our dosing notation $C$ to create a potential outcome $C^{(A)}$ for correct dosing under each treatment assignment. This allows us to classify participants into groups based on their potential to receive the two doses correctly after being assigned to each treatment. This creates four potential types (or strata) of participants: Participants that would not receive two doses in the correct time window if assigned to either treatment, participants that would receive two doses in the correct time window if assigned to one treatment, but not the other, and finally, participants that would have received two doses in the correct time window if assigned to either treatment. The last type are the participants forming the principal stratum of interest.

With the notation extended to create the classifications we can write down the second causal estimand.
$$
{100 \cdot \left( 1 - \frac{\mathbb{E}[Y^{\:(a\:=\:1)} \: \vert \: C^{(a\:=\:0)} = C^{(a\:=\:1)} = 1 ]}{\mathbb{E}[Y^{\:(a\:=\:0)} \: \vert \: C^{(a\:=\:0)} = C^{(a\:=\:1)} = 1 ]} \right) }
$$
In words, this effect is the expected percentage reduction in risk of COVID-19 infection for participants assigned mRNA-1273 vaccine or placebo that are members of the stratum that would receive both doses of their assigned treatment regardless of assignment. We also note that the definition of this effect is in terms of potential outcomes that cannot both be observed. Therefore, it is only possible to estimate this effect using strong and untestable assumptions. Currently within the literature there are differing opinions of the utility of the principal stratum approach \citep{Stensrud2022, Bornkamp2021}.

This example contrasts the difference between effects that apply to the entire population (estimand 3.1) and effects that apply to a reduced number of participants from the original population (estimand 3.2).

\section{Discussion} \label{sec:discussion}
\subsection{Comparing the ICH and Causal Paradigms}

The examples above demonstrate how both the estimand framework and causal inference methodology can be used to define estimands. Both approaches allow clinical teams flexibility in defining the specific effect of interest and facilitate the discussion about what overall effect is being considered in light of various complicating factors. The ICH estimand is explicit with five components that must be defined, whereas the causal estimand is the result of defining the ‘treatment’ variables and potential outcomes to be compared and the assumed causal structures impacting the outcome (often via a DAG).
For causal estimands, factors such as treatment assignment, stopping treatment, and using additional or rescue medication are specified in the algebra which contains a considerable amount of information in a very subtle way. It is precise, but the concepts it relies on, and subtleties of specification may be harder to comprehend. This mathematical expression, together with a well-defined population and detail on the outcome of interest, provides the same overall information as an ICH estimand. Further, the use of DAGs allows the assumed causal structure of the outcome, assignment, and other factors to be communicated to guide the estimation of the effect targeted.
In contrast, the ICH estimand is not mathematical and is expressed either in a formal table or a single dense paragraph of text. This is likely to be more accessible to clinical trialists without formal causal and mathematical training. However, because the text is not as precise as the causal algebra, it is important to define the ICH estimand using detailed statements.
As demonstrated in our examples, the concept of an intercurrent event (IE) and strategies to address them borrow ideas and concepts from causal inference. It is reasonably intuitive to consider the concept of an IE to be the same as post-baseline mediator or time varying treatment developed as part of the causal inference literature. 

A number of authors have previously described potential linkages between the types of causal effects and IE handling strategies, but no consensus has emerged \citep{Lipkovich2020, Scharfstein2017, Han2023}.  In this paper we have explored a number of examples to show the connection between IE handling and causal inference.

In hindsight, given the causal inference and biostatistics literature overlap heavily, and the term estimand has existed for many years, it might have been helpful for ICH E9 (R1) to have discussed linkages to causal inference terms and concepts explicitly. We think the creation of the ICH estimands framework was needed to provide clarity on what is estimated in clinical trials, and all five components are required to gain the understanding of what is being estimated, but the lack of discussion linking it to established causal concepts has created some confusion.

Finally, the examples demonstrate that what is called an estimand in the two paradigms may only partially overlap but are essentially both targeting the same thing: a population-based summary of a causal effect on an outcome in a particular population. 

\subsection{General Comments on Causal Inference}

When applying causal inference to clinical trials in drug development, only two treatment effects of interest have commonly been identified: (i) the effect of assignment to the intervention and (ii) the effect of adhering to the intervention as specified in the trial protocol \citep{Sterne2019, Hernan2017, Cochrane2019}. The latter is often described as the per protocol effect, specifically \textit{“... the causal effect of treatment that would have been observed if all individuals had adhered to their assigned treatment as specified in the protocol of the experiment.”} \citep{Hernan2024, Hernan2017}. The term "per-protocol effect" may be confusing to non-statisticians in drug development since it is often interpreted incorrectly as an analysis of a per-protocol population \citep{Keene2023}. Similarly, the term "adherence" is sometimes ill-defined and can refer to continuing to take a medication or to some formal measure of compliance. It is also unclear in the causal inference literature how events such as use of rescue medications or surgery are accounted for when discussing adherence, particularly in trials where these additional interventions are indicated in the protocol.
While “assignment” or “adherence” effects may answer important scientific questions, there are other research questions that a multi-stakeholder trial would regularly aim to answer and contrary to this perceived dichotomy, causal inference allows the definition of effects that are not purely from assignment or perfect protocol compliance. Therefore, suggestions that the assignment and adherence effects are the only important questions is unhelpful \citep{Keene2023}.

\subsection{General Comments on ICH E9 (R1)}

Although finalized in 2019, ICH E9(R1) still contains some ambiguity on certain points. As a result, further work has been published suggesting ICH estimands should have certain additional properties \citep{Mtze2024}. We comment on some aspects of the addendum in general below and would welcome further discussion with regulators. 

In general, the strategy labelled “treatment policy” is not particularly accurate when used for handling discontinuation of assigned treatments. In many diseases, when the randomized treatment fails, the actual treatments participants receive would be at the discretion of the physician and would typically contain new treatments or no treatment at all. In these cases, there is not a well-defined “treatment policy” and the strategy itself would be better labelled as the assessing the effect of assignment.

A further source of ambiguity relates to the explanation of the principal stratum strategy.  As written, it has the potential to mislead the reader into believing the stratum is a post baseline subgroup of the patients from each randomized group that are observed to satisfy the IE constraint, rather than the effect for patients that would have satisfied the constraint if they were assigned to either treatment. We included the COVID example to illustrate which patients a principal stratum strategy effect can correspond to. There are also contrasting opinions on the clinical relevance of this strategy in general \citep{Stensrud2022, Bornkamp2021}. 

Although the hypothetical strategy can be viewed as the effect if the IE had not occurred \citep{Fletcher2022}, non-statisticians sometimes find the term “hypothetical” problematic. This is because the term suggests something is being estimated that is different from factual evidence and, for this reason, the strategy is increasingly viewed as irrelevant to use of the medicine in clinical practice. However, applying this strategy to IEs, attempts in many cases to assess the causal effect of the medicine itself and this effect may be relevant to patients and prescribers. An effect described in the product description (label) for a medication based on a treatment policy approach will reflect a mixture of results from participants taking a range of possible treatments.  Clear identification of the causal effect of the treatment itself would be helpful for patients that are considering whether to take the medication.

A potential approach to addressing this is the attributable estimand described by Darken et al. \citep{Darken2020}.  This estimand uses different strategies for different intercurrent events. Those events that are considered to be adversely related to randomised treatment (e.g. discontinuation of treatment due to adverse events or lack of efficacy) are considered attributable and handled with a composite strategy (hence become part of the endpoint of interest), while a hypothetical strategy is used for intercurrent events not considered to be related to randomised treatment (e.g. treatment discontinuation due to administrative reasons such as loss to follow-up).  Estimates of treatment effect from an attributable estimand may be of more use to patients and prescribers than the common practice of providing estimates based on a treatment policy strategy.

\section{Conclusion} \label{sec:conclusion}
We have discussed estimands as defined in ICH E9 (R1) as part of the “estimand framework” and causal inference field. We provided a brief commentary on where both concepts of an estimand originated and used real trial examples with clinical questions of interest to illustrate how both ICH and causal estimands could be formulated. We then compared the similarities and differences between the approaches.
We conclude both can be used to quantify a population-based summary of a target effect on an outcome for a given population. Both paradigms rely on assumptions that underpin causal inference, and both have their use in the field of clinical trials. The ICH estimand is a description-based definition which is likely to be more accessible to non-mathematicians. The causal estimand is more precise due to its mathematical nature but may be harder to grasp for people without the causal training to mathematically define the target of inference and communicate their assumptions through causal graph diagrams.
Although we have presented both interpretations of estimands side by side for understanding and comparison, we would encourage practitioners to consider them as complementary and not competing frameworks. Both approaches have attractive features and combining them allows the trialist more options for communicating what is being estimated in clinical trials. We encourage people familiar with the estimands framework to use causal inference and vice versa.

\section{Acknowledgement} \label{sec:acknowledgement}
The authors would like to thank Matt Psioda and Ben Hartley for discussions on potential outcomes and composite IE handling strategies. 

\section{Declaration} \label{sec:declaration}
TD owns shares in GlaxoSmithKline. JWB and his employers have received fees for statistical consultancy from AstraZeneca, Bayer, Novartis, and Roche. DW owns shares in AstraZeneca. ONK owns shares in GlaxoSmithKline.


\bibliography{wileyNJD-AMA}

\end{document}